\documentclass[sigconf]{acmart}

\AtBeginDocument{%
  \providecommand\BibTeX{{%
    \normalfont B\kern-0.5em{\scshape i\kern-0.25em b}\kern-0.8em\TeX}}}

\setcopyright{acmcopyright}
\copyrightyear{2024}
\acmYear{2024}
\acmDOI{XXXXXXX.XXXXXXX}

\acmConference[WWW ’24 Companion]{Companion Proceedings of the ACM Web Conference 2024}{May 13--17,2024}{Singapore}
\acmPrice{15.00}
\acmISBN{978-1-4503-XXXX-X/18/06}

\usepackage{graphicx}
\usepackage{subcaption}
\usepackage{multirow}
\usepackage{stfloats}
\begin{document}

\title{Query-LIFE: Query-aware Language Image Fusion Embedding for E-Commerce Relevance}
\author{Hai Zhu}
\authornote{This work was done when Hai Zhu was an intern at Lazada.}

\orcid{1234-5678-9012}

\affiliation{%
  \institution{University of Science and Technology of China}
  \city{Hefei}
  \country{China}
  \postcode{43017-6221}
}
\email{SA21218029@mail.ustc.edu.cn}

\author{Yuankai Guo, Ronggang Dou,  Kai Liu}
\authornote{Corresponding author.}
\affiliation{%
  \institution{Lazada}
  \city{Hangzhou}
  \country{China}}
\email{{guoyuankai.gyk,ronggang.drg,baiyang.lk}@alibaba-inc.com}


\renewcommand{\shortauthors}{Hai Zhu, et al.}

\begin{abstract}

Relevance module plays a fundamental role in e-commerce search as they are responsible for selecting relevant products from thousands of items based on user queries, thereby enhancing users experience and efficiency. The traditional approach models the relevance based product titles and queries, but the information in titles alone maybe insufficient to describe the products completely. A more general optimization approach is to further leverage product image information. In recent years, vision-language pre-training models have achieved impressive results in many scenarios, which leverage contrastive learning to map both textual and visual features into a joint embedding space. In e-commerce, a common practice is to fine-tune on the pre-trained model based on e-commerce data. However, the performance is sub-optimal because the vision-language pre-training models lack of alignment specifically designed for queries. In this paper, we propose a method called \textbf{Query-LIFE} (\textbf{Q}uery-aware \textbf{L}anguage \textbf{I}mage \textbf{F}usion \textbf{E}mbedding) to address these challenges. Query-LIFE utilizes a query-based multimodal fusion to effectively incorporate the image and title based on the product types. Additionally, it employs query-aware modal alignment to enhance the accuracy of the comprehensive representation of products. Furthermore, we design GenFilt, which utilizes the generation capability of large models to filter out false negative samples and further improve the overall performance of the contrastive learning task in the model. Experiments have demonstrated that Query-LIFE outperforms existing baselines. We have conducted ablation studies and human evaluations to validate the effectiveness of each module within Query-LIFE. Moreover, Query-LIFE has been deployed on Miravia Search\footnote{Miravia is a local-to-local e-commerce platform in Spain incubated by Lazada, as one part of Alibaba International Digital Commerce (AIDC) Group. \url{https://www.miravia.es/}}, resulting in improved both relevance and conversion efficiency.

\end{abstract}




\maketitle

\section{Introduction}
Nowadays, since Internet penetration is increased to a high level, online shopping has become to a highly convenient manner for consumers. Everyday, millions of users search \& browse products, and maybe finally place orders in e-commerce platforms. Consequently, the relevance of products exposed to users triggered by their queries plays a crucial role in users shopping experiences, and also the transaction efficiency. Therefore, it is essential to accurately judge whether the candidate products are relevant to the user intentions for an e-commerce search engine.

Traditionally, relevance model~\cite{bm25,dssm,chang2021extreme,arc,yao2021learning} have primarily relied on textual information, such as query and product descriptions (title, attribute, etc) to judge the relevance between queries and products. However the product information also includes images which captures a significant portion of user attention during browsing products, thus it is becoming increasingly essential to incorporate image into relevance modeling. This integration of both image and text data has the potential to provide a more comprehensive understanding of the products and better capture user intent.

\begin{figure*}[htb]
\includegraphics[width=0.8\textwidth]{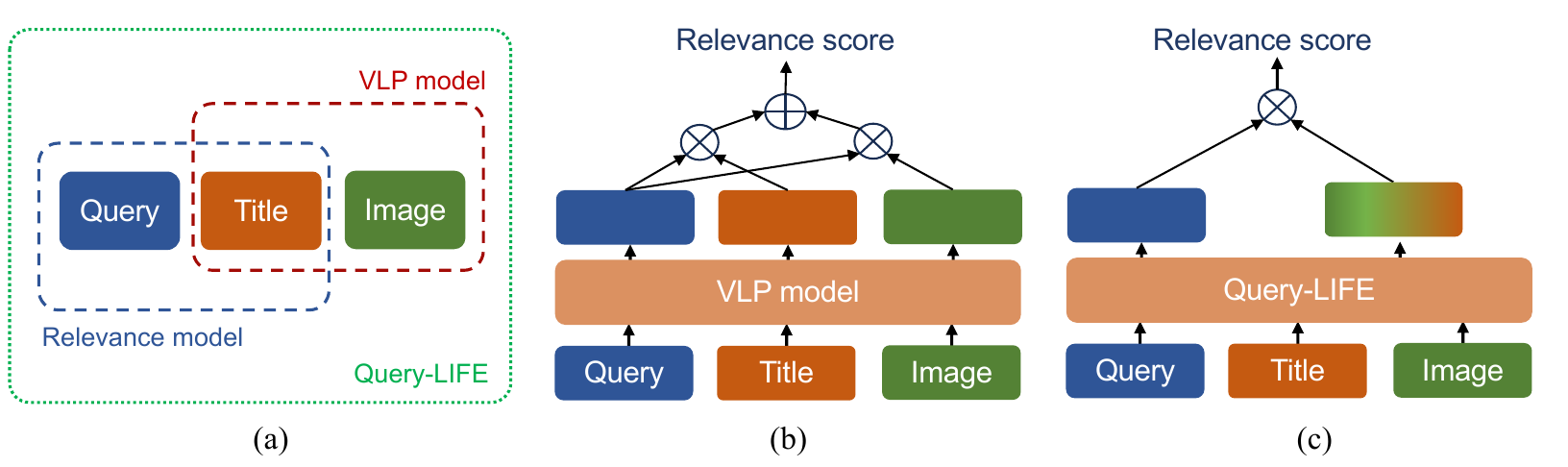}
  \caption{(a) The relationship of relevance model, VLP model and Query-LIFE. (b) VLP model divide-and-conquer approach for relevance task. (c) Query-LIFE fusion approach for relevance task.}
  \Description{.}
  \label{fig:relation}
\end{figure*}

Some core information may be missing from product titles, just like the cases in Table ~\ref{tab:online_case}. In these cases, if only based on the product titles, it is hard to accurately match the relevant product with search query. However, product images can provide incremental valuable information for relevance judgement. In recent years, there has been a surge of vision-language pre-training models (VLP)~\cite{blip2,albef,align,beit,wang2021simvlm}. As shown in Figure~\ref{fig:relation}(a), these VLP models typically consist of both text and image encoders, and leverage text-image contrastive learning to align representations across different modalities. They have demonstrated remarkable performance in various general tasks, such as image captioning, visual question answering, and text-image retrieval. It is worth noting that these VLP models can extract image features to enhance the representation of products with ambiguous titles, or correct the representation of products with misleading titles in e-commerce relevance task.   Current VLP models in e-commerce relevance task often employ a divide-and-conquer approach. As shown in Figure~\ref{fig:relation}(b), they first conduct VLP model extracts query, title, and image representations, and then add query-title similarity and query-image similarity as the final relevance score.  However, different types of products contain varying weights of information in images and titles. For example, electronic products often have important parameters listed in their titles, while clothing items rely more on visual elements depicted in the images, such as designed style, texture, material, color, etc.

\begin{table}[htb]
  \centering
  \caption{Both product images and titles can be incorporated together to judge the search relevance with queries.}
   \label{tab:online_case}
  \begin{tabular}{  p{2.5cm} | p{1.5cm} | p{3cm}  }
    \toprule
    Query & Image & Title   \\ \midrule
    men's winter coat
    & \begin{minipage}[b]{0.15\columnwidth}
		\centering
		\raisebox{-.5\height}{\includegraphics[width=\linewidth]{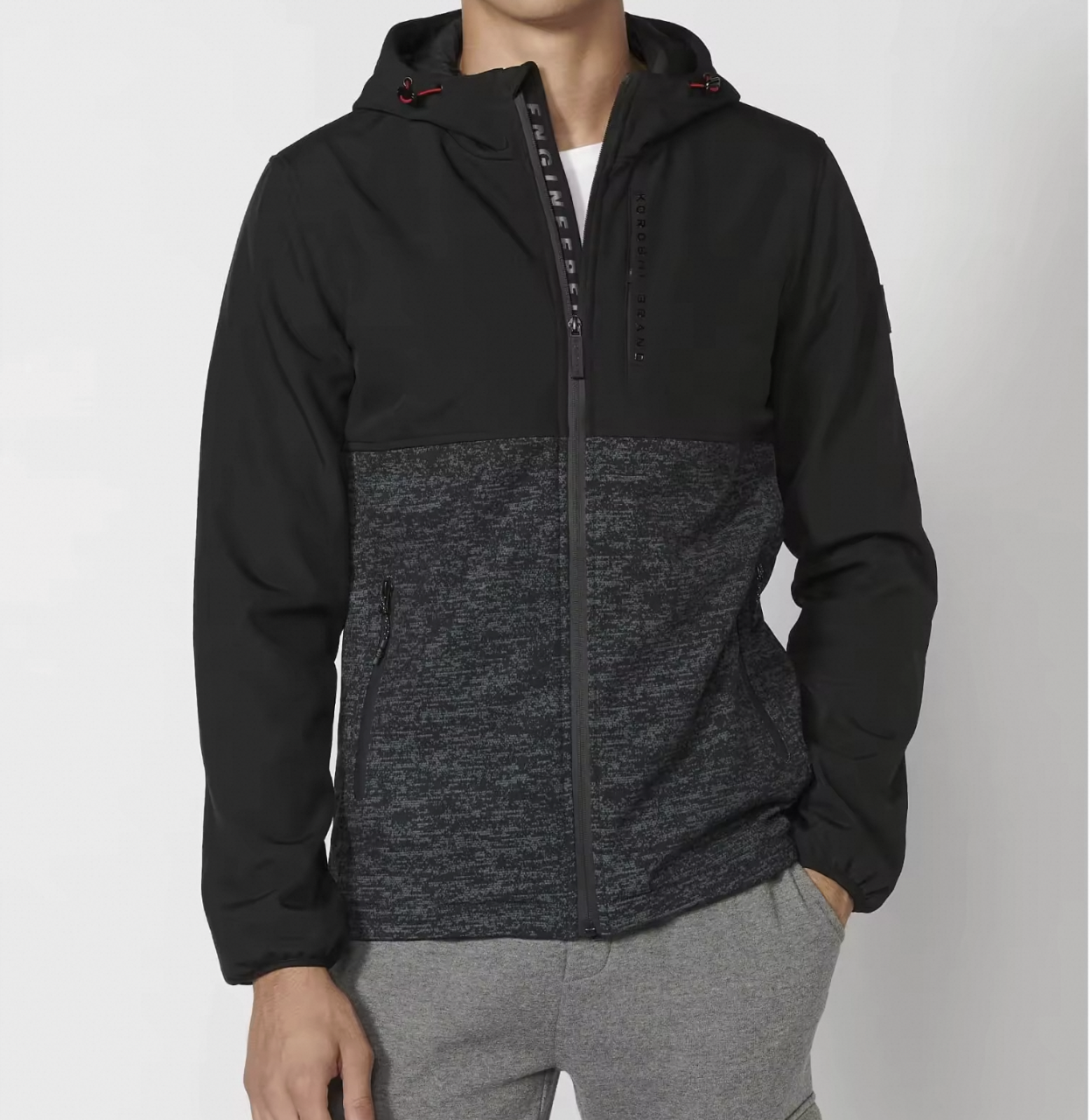}}
	\end{minipage}
    & Koroshi Jacket in two colors, water-repellent, with hood, for Men
    \\ \midrule
    air-conditioning
    & \begin{minipage}[b]{0.15\columnwidth}
		\centering
		\raisebox{-.5\height}{\includegraphics[width=\linewidth]{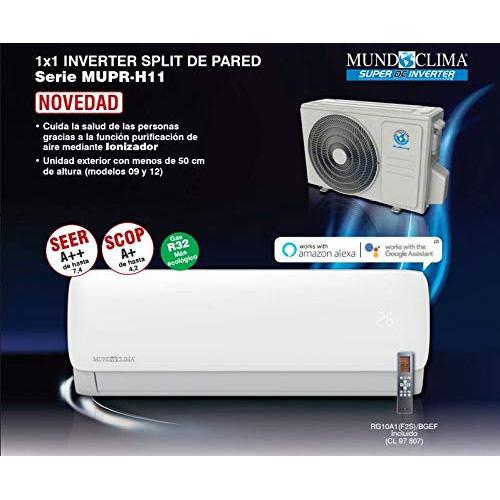}}
	\end{minipage}
    & Split 1x1 MUNDOCLIMA MUPR12 H11 3027frig R32
    \\ \midrule
    golden necklace
    & \begin{minipage}[b]{0.15\columnwidth}
		\centering
		\raisebox{-.5\height}{\includegraphics[width=\linewidth]{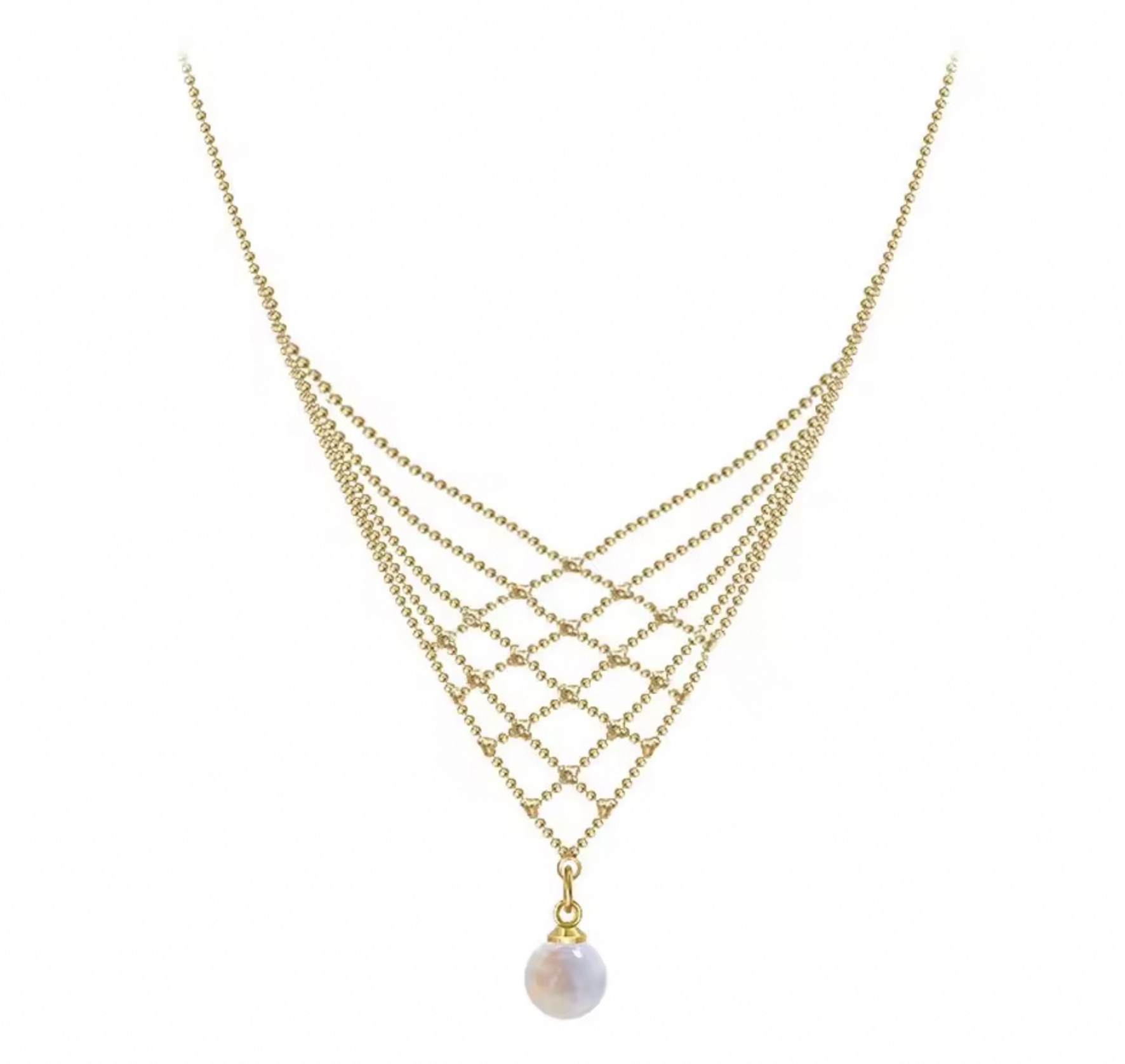}}
	\end{minipage}
    & Elegant necklace with col-layered pearl gent
    \\\bottomrule
  \end{tabular}
\end{table}

In this paper,  we  propose a novel approach called \textbf{Q}uery-aware \textbf{L}anguage \textbf{I}mage \textbf{F}usion \textbf{E}mbedding(referred as \textbf{Query-LIFE}) for e-commerce relevance modeling. As shown in Figure~\ref{fig:relation}(a), it integrates query, title, and image into relevance  tasks. Firstly, we randomly sample <query,title,image> triplet data from the online user behavior logs as pre-training data. Secondly, unlike the divide-and-conquer approach, we propose a dynamic fusion as multi-modal representation of product. Hence, we use inner product of <query, multi-modal
representation> to measure the relevance between one query and the multi-modal representation above, as shown in Figure ~\ref{fig:relation}(c). Thirdly, we use supervised contrastive learning and utilize generating ability both from multi-modal large model and large language model to filter out the false negative samples. Finally, we fine-tune the proposed model using the manually annotated triplet data. Our contributions can be summarized as follows:
\begin{itemize}
    \item We propose query-aware multi-modal fusion tailored for e-commerce relevance task. Instead of divide-and-conquer approach,  Query-LIFE proposes multi-modal representation and adopt <query, multi-modal representation> similarity as the relevance score.  It effectively generates dynamic fusion representations that incorporate product images and text based on the product types.
    
    \item We also propose query-based modal alignment module utilizes supervised contrastive learning to align the multi-modal representation of products guided by the search query. improves the representation quality and enhances the matching between user queries and dynamic fusion representations of products.

    \item Generating and Filtering (GenFilt) is designed to mitigate the impact of false negative samples during the training process. By generating additional positive samples and filtering out false negative ones, it helps to improve the quality of the training data, leading to enhanced model performance and robustness.
    
    \item We carried out extensive experiments on both offline and online A/B experiments, which have shown that it outperforms existing VLP models and relevance models in e-commerce relevance task. Our model has been successfully deployed in Miravia Search.
\end{itemize}

\section{Related Work}
\subsection{Vision-Language Pre-training}
The emergence of pre-training models, such as BERT~\cite{bert}, GPT3~\cite{brown2020language}, and ViT~\cite{vit}, has led to significant advancements in NLP and CV tasks, achieving state-of-the-art results. Recently, researchers have extended the pre-training approach to the vision-language (VL) domain, resulting in the development of several impressive VL models (e.g. CLIP~\cite{clip} and ALIGN~\cite{align}). These VLP models have demonstrated impressive performance in various multi-modal downstream tasks, such as image captioning, vision question answering, and cross-modal retrieval. They achieve this by leveraging large-scale image-text pairs and then employing contrastive learning to align images and text in the joint embedding space. These VLP models are divided into two categories: object-detector (OD)-based VLP models (e.g., UNITER~\cite{chen2020uniter}, OSCAR~\cite{li2020oscar}) and end-to-end VLP models (e.g., ALBEF~\cite{albef},BLIP~\cite{blip2})). OD-based VLP models rely on bounding box annotations during pre-training and require high-resolution images for inference, making them both annotation-expensive and computationally demanding. In contrast, end-to-end VLP models directly utilize image patch features as input to a pre-trained ViT model. This eliminates the need for costly annotations and significantly improves inference speed. As a result, end-to-end VLP models have gained traction in recent research~\cite{chen2021kb,kim2021vilt}. Therefore, we also adopt the end-to-end VLP model in this paper.

\subsection{E-commerce VLP Model}
There are also  some VLP models specifically targeted at e-commerce scenarios.  FashionBERT~\cite{gao2020fashionbert} was the first vision-language pre-train model which adopts mask language loss and title-image contrastive learning.  Later on, Kaleido-BERT~\cite{zhuge2021kaleido} further adopts several self-supervised tasks at different scales to focus more on title-image coherence. EI-CLIP~\cite{ma2022eiclip} proposed an intervention-based entity-aware contrastive learning framework. KG-FLIP~\cite{jia2023kgflip} proposed a knowledge-guided fashion-domain language-image pre-training framework  and utilizes external knowledge to improve the pre-training efficiency. 
MAKE~\cite{zheng2023make} introduces query encoder and propose modal adaptation  and keyword enhancement modules to  improve text-to-multimodal matching. However, these e-commerce VLP models focus on multi-modal retrieval rather than relevance task, so their losses are designed for retrieval. On the contrary, our work focuses on relevance task and our motivation is to deeply integrate the VLP model to enhance the product's relevance score in e-commerce scenarios.

\section{Method}
\subsection{Model Architecture}
\begin{figure*}
  \includegraphics[width=\textwidth]{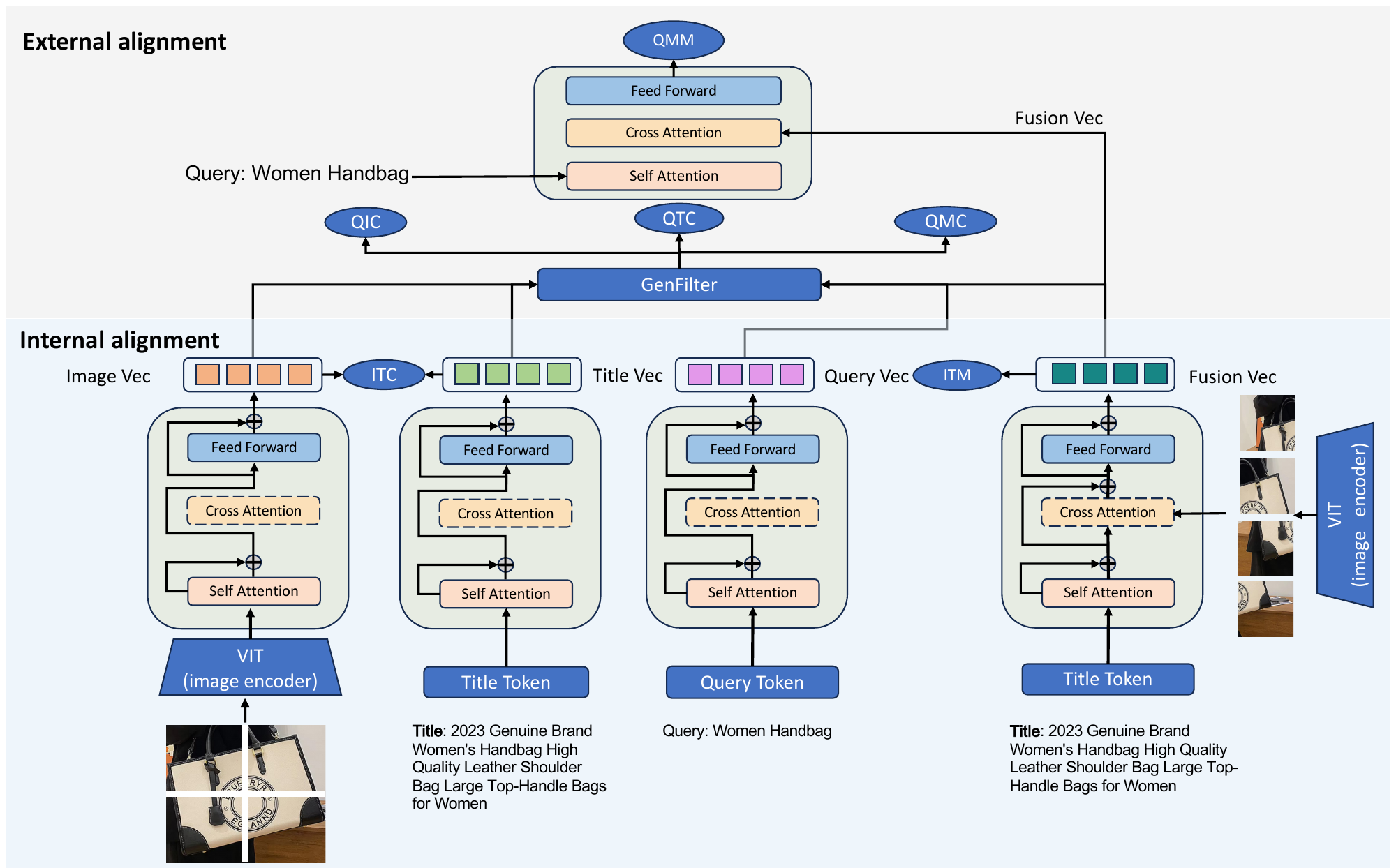}
  \caption{Overview of Query-LIFE.  The overall training process is divided into internal aligment and external alignment.}
  \Description{.}
  \label{fig:overreview}
\end{figure*}

In this section, we will  introduce our model architecture in detail. As shown in  Figure~\ref{fig:overreview}, The entire model training is divided into internal alignment and external alignment. Internal alignment is used to align the features of product titles and images. External alignment is used to align the relevance  between user queries and products.   The model architecture consists of three modules: an image preprocessing backbone, a universal modal encoder and GenFilt. Visual transformer (ViT)~\cite{vit} is deployed as our image preprocessing backbone, which divides the image into patches and encodes them as a sequence of embeddings with an additional $[CLS]$ token to represent the global image features. The universal modal encoder includes self-attention layer,  cross-attention layer and  feed-forward layer. It extracts different modalities features through the interaction of different layers. For the text modal, the input undergoes tokenization and then interacts with the self-attention layer and the feed-forward layer. Similarly, for the image modal, the input image is first processed by the ViT and   follows the same process as the text modal. In the case of multimodal features, the text modal interacts with the image modal through the cross-attention layer after the self-attention layer.  GenFilt is designed for filtering false negative sample during in-batch sampling. 

Inspired by the relevance learning framework proposed by Jiang et al~\cite{jiang2019unified},  we also propose a three-stage training framework. In the first stage, we leverage a large set of products title-image pairs for contrastive learning, which is pre-trained for internal alignment of products in section 3.2.  In the second stage,  we sample million <query,title,image> positive pairs from the online clicking log of Miravia Search, then pre-train for external alignent between user queries and products  in the section 3.3 and 3.4. In the third stage, we utilize manually labeled <query, title, image> triplet data  to fine-tune the alignment between products and user queries furtherly. 
 
\subsection{Vision-Language Pre-training}
The VLP model utilizes the Image-Text Contrastive (ITC) loss to align the image features and the text features, which makes positive image-text pairs have similar representations and reduces the similarity between negative pairs~\cite{radford2021learning,albef}. The ITC loss has proven to be an effective objective for enhancing vision and language representation even in the absence of labeled data. The formula is as follows:
\begin{equation}
    \mathcal{L_{ITC}} = -\frac{1}{N}\sum_{i=1}^{N}log\frac{exp(Z_{T_i} \cdot Z_{I_i}/\tau)}{\sum_{j=1}^{N} exp(Z_{T_j} \cdot Z_{I_j}/\tau)}.
\end{equation}
where $Z_T$ and $Z_I$ are normalized text and image embeddings, $Z_{I_i}$ is the $i$-th positive image sample in the batch, and $N$ and $\tau$ batch size and temperature parameter respectively.

\subsection{Query-based Modal Alignment}
In e-commerce search scenarios, the relevance of products is highly dependent on the user queries. However, user queries are short and brief, while sellers often add redundant keywords in titles to hack the search engine indexing. Calculating relevance solely based on query and title can easily result in errors in relevance scoring. If misleading words or insufficient information are present in the title, this problem becomes even more severe.

Obviously, there is an imbalance problem between query and product information in e-commerce relevance task.  To mitigate the impact of above problem on relevance scoring, we introduce image information to improve product representation. In addition, we introduce the concept of title-image fusion representation for products (referred as multi-modal representation or $\mathcal{M}$ representation). $\mathcal{M}$ representation is defined as the interaction between the product title and the image. Unlike divide-and-conquer approach, we adopt universal modal encoder to  represent $\mathcal{M}$  and use inner product to calculate the relevance between the item and query. To further align the $\mathcal{M}$ representation with user queries, we adopt the query-$\mathcal{M}$ contrastive (\textbf{QMC}) loss. Additionally, we incorporate the query-title contrastive (\textbf{QTC}) loss to align the query with title. Simultaneously, the query-image contrastive (\textbf{QIC}) loss is utilized to further align the product image with the query. These loss functions play a crucial role in aligning user queries and different product  modalities and enhancing the relevance scoring.

Unsupervised contrastive learning utilizes large amounts of unlabeled data to increase the similarity between positive samples while decrease the similarity between negative samples. However, in e-commerce relevance task,  the same query often generates positive pairs with different products.  Compared with triplet loss and unsupervised contrastive learning, supervised contrastive learning introduces labeled negative samples and  can contain more positive samples in a mini-batch, which is more suitable for relevance task. The formula is defined as follows:
\begin{equation}
    \mathcal{L} = -\frac{1}{N}\sum_{i=1}^N\left\{\frac{1}{|P(i)|}\sum_{p\in P(i)}\left[log\frac{exp(Q_i \cdot Z^x_{p}/\tau)}{{\sum_{j=1}^N}  exp(Q_i \cdot Z^x_{j}/\tau)}\right]\right\}.
\end{equation}
where $P(i)$ is all positive samples in the $i$ batch, $Q$ is normalized query embedding and $Z^x_{p}, x\in[I,T,M]$, $p\in P(i)$, are normalized image/text/$\mathcal{M}$ modal embedding in positive samples. $\tau$ and $N$ are temperature parameter and batch size.  Bringing in different modal by $x\in[I,T,M]$, this loss function can represent QIC, QTC and QMC loss respectively.

\subsection{Query-based Modal Fusion}
We utilize image-text matching (ITM) to learn the $\mathcal{M}$ representation of the product. The objective of ITM is to learn a title-image fusion representation that captures the alignment between the image and text modalities. In ITM, we frame the task as a binary classification problem where the model predicts whether an image-text pair is positive or negative. To obtain the matching score, we pass the model's output through a two-class linear classifier, yielding a logit. We employ a hard negative mining strategy~\cite{align} and leverage labeled data. Hard negative mining strategy samples negative pairs with higher contrastive similarity within a batch. Consequently, these informative negative pairs contribute to better align between the image and text modalities. The ITM loss can be expressed as:
\begin{equation}
\label{itm}
    \mathcal{L}_{ITM} = -E_{(I,T) \sim P}[log\{P(y_{(I,T)}|(I,T)\}]
\end{equation}
where $P$ is a distribution of in-batch samples, $y_{(I,T)} \in (0,1)$ represents whether the image $I$ and the text $T$ are matched, and $P(y_{(I,T)}|(I,T))$ is the  output of the  multi-modal embedding followed by a two-class linear classifier.

We acknowledge that image-text matching alone may not be sufficient, as different types of products contain varying amounts of information in their images and titles. For example, electronic products often have important parameters listed in their titles, while clothing items rely more on visual attributes such as material, color, and size displayed in the images. To enable the model to learn a more effective fusion representation, we introduce query-$\mathcal{M}$ matching (\textbf{QMM}). QMM not only allows the model to extract features from both the images and titles, but also giving different weights to each modality based on the user queries. This enables the model to generate fused representations with a query-aware bias. QMM and ITM share the same loss function listed in Eq.~\ref{itm}.  Finally, the Query-LIFE model loss contains vision-language pre-training (VLP), query-based modal alignment (QMA) and query-based modal fusion (QMF):
\begin{equation}
    \mathcal{L}_{total} = \underbrace{\mathcal{L}_{ITC}}_{VLP}+\underbrace{\mathcal{L}_{ITM}+\mathcal{L}_{QMM}}_{QMF}+\underbrace{\mathcal{L}_{QIC}+\mathcal{L}_{QTC}+\mathcal{L}_{QMC}}_{QMA}.
\end{equation}

\subsection{GenFilt}
Most VLP models adopt  in-batch sampling  to generate image-title negative pairs. However, in the <query, title, image> triplet data, multiple user queries may be relevant to the multiple products.  In-batch sampling will introduce false negative samples. Those similar even same queries are mistakenly treated as negative samples and thus compromise relevance score.
\begin{figure*}[htb]
  \includegraphics[width=0.90\textwidth]{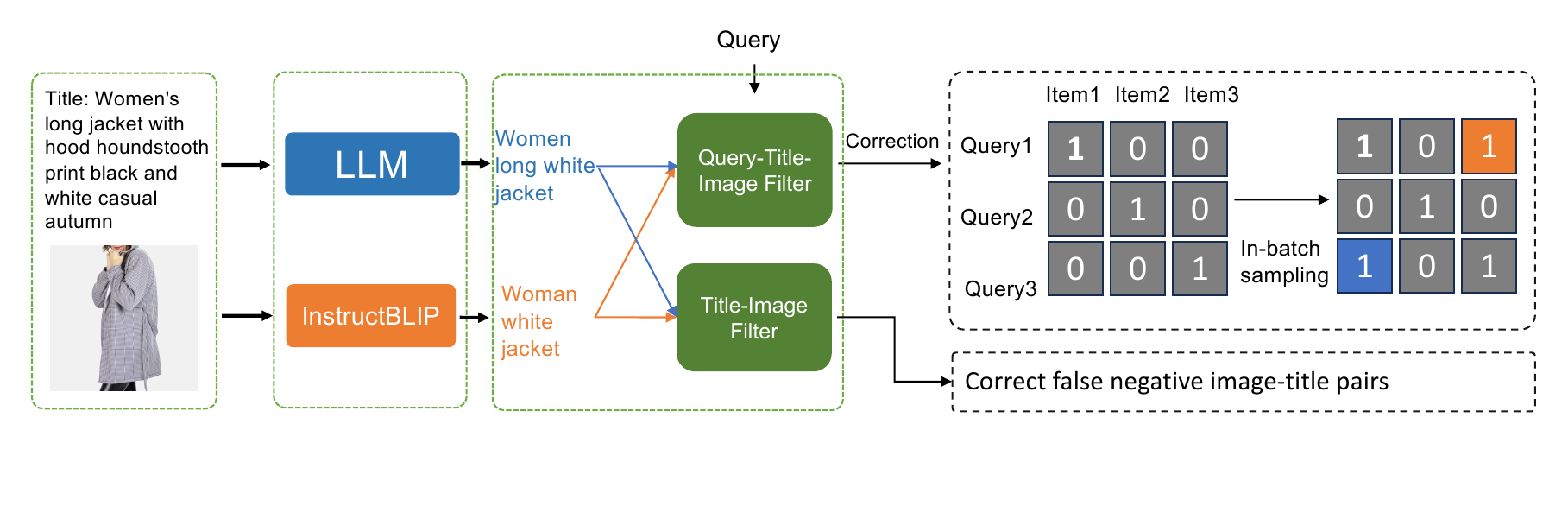}
  \caption{Overview of  GenFilt. GenFilt adopts LLM and InstructBLIP for feature generation. Then compare the similarity of  query-product pairs and correct the false negative pairs. In addition, GenFilt can also calculate the similarity of image-title pairs and correct false negative pairs.}
  \Description{.}
  \label{fig:correction}
\end{figure*}

Inspired by CapFilt~\cite{blip2}, we propose a method called Generating and Filtering (GenFilt) to  address the impact of false negative samples on the training process. It enhances the quality of the training data using large model generation capabilities. As illustrated in Figure~\ref{fig:correction}, GenFilt consists of two modules. The first module is generating, we employ a large language model (LLM) and a multi-modal model (InstructBLIP)~\cite{dai2023instructblip} to extract key text features from the product title and image respectively. The second module is filtering, we calculate the similarity between  image feature and  text feature (I-T), the similarity between  query feature and  image feature (Q-I) and the similarity between  query feature and  text feature (Q-T). Finally, we set a threshold $\sigma$ based on these similarities, and  the similarity of query-product pairs ((Q-I+Q-T)/2) and image-text pairs (I-T) above the threshold are also corrected as positive samples.

\section{Experiments}
\subsection{Baselines and Datasets}
\textbf{Large-scale Industrial Datasets.} There are three training datasets. The first dataset samples 5M product title-image pairs. The second dataset samples 1.3M <query,title,image> positive pairs from the online clicking log of Miravia Search.  The third dataset  has  200,000 <query,title,image> labeled data, and 30K samples are selected as the evaluation set, where the ratio of positive to negative is 1:1.

\textbf{Baselines.} In our experiments, we compare  Query-LIFE with several strong baselines, including BERT~\cite{bert}, CLIP-zeroshot~\cite{clip} and  BLIP2-zeroshot and BLIP2-FT(finetune)~\cite{blip2}. CLIP and BLIP2 are two-tower models, while BERT is a single-tower and text-modal relevance model. Only the <query, title> pairs from the <query, title, image> triplet data are used for training. This approach does not incorporate image information. These baselines have shown impressive performance in their respective domains.  For the CLIP and BLIP2, we concatenate the query and title as text information and train the models with images. To ensure fair comparison,  all models are trained on the 200,000 <query, title, image> triplet data samples.

\subsection{Experiment Implementation}
We select the state-of-the-art pre-trained vision transformer ViT-g/14 from EVA-CLIP~\cite{fang2023eva} and frozen it. We  remove the last layer of the ViT and use the second last layer’s output features, which  leads to slightly better performance and is the same as BLIP2~\cite{blip2} setting. The universal modal encoder is composed with 12 layers of transformers, each layer contains self-attention layer,cross-attention layer and feed forward layer. We initialize the universal modal encoder with the pre-trained
weights of BERT-base~\cite{bert}, whereas the cross-attention layers are randomly initialized. In total, universal modal encoder contains 188M parameters. We train the model for 10 epochs with a batch size of 512 on 16 NVIDIA A10 GPUs. We utilize the AdamW optimizer with $\beta_1 = 0.9, \beta_2 = 0.98$, and a weight decay rate of 0.05. For learning rate scheduling, we employ a cosine decay strategy with a maximum learning rate of $1e^{-4}$ and a linear warmup of 2k steps. GenFilt threshold $\sigma$ is set to 0.9.

\subsection{Evaluation Metrics}
\textbf{Offline Evaluation Metrics.} We  consider manual annotation as the ground truth, where relevance is indicated by the labels of 1 (relevant) and 0 (irrelevant). This task can be treated as a classification problem. In e-commerce scenarios, the Area Under Curve (AUC) is commonly used as the evaluation metric~\cite{yao2021learning}.  Additionally, we utilize the Precision-Recall curve for evaluation. This curve represents the trade-off between precision and recall, with recall as the x-axis and precision as the y-axis. In addition,  we employ Recall@K (R@K) as the metric. Recall@K is widely used in search and recommendation systems. To calculate Recall@K, we randomly select 10K unique <query,title,image> triplet data from the online clicking log of Miravia Search. For each query, we consider the ground-truth product as well as 100 other randomly sampled products as the candidate rank set. We calculate the similarity between the query $\rightarrow$ title, query $\rightarrow$ image, and query $\rightarrow$ $\mathcal{M}$ and sort the candidate rank set based on these similarities. Recall@K measures the percentage of ground-truth matches that appear in the top-K ranked list~\cite{gao2020fashionbert}.

\begin{table*}[htb]
\centering
\caption{Offline results compared with different baselines.}
  \label{tab:baseline}
\begin{tabular}{ccccccccccccc}
\toprule
\multirow{2}{*}{}    & \multicolumn{4}{c}{Query $\rightarrow$ Title} & \multicolumn{4}{c}{Query $\rightarrow$ Image} & \multicolumn{4}{c}{Query $\rightarrow$ $\mathcal{M}$} \\\cmidrule(lr){2-5}\cmidrule(lr){6-9}\cmidrule(lr){10-13}
                     & R@5    & R@10   & R@20   & AUC  & R@5    & R@10   & R@20   & AUC  & R@5    & R@10   & R@20  & AUC   \\\midrule
BERT              & \textbf{0.142}       & {0.186}     & \textbf{0.351}     & \textbf{0.871}     & -      & -      & -     & -     & -       & -     & -     & -     \\
CLIP-Zeroshot     & 0.068       & 0.125     & 0.272     & 0.542     & 0.068      & 0.147      & 0.272     & 0.554     & -       & -     & -     &-    \\
BLIP2-Zeroshot & 0.134       & 0.204     & 0.318     & 0.565     & \textbf{0.090}      & 0.159      & 0.272     & 0.575     & -       & -     & -     &- \\
BLIP2-FT     &0.113     &0.170    &0.272    & 0.752  &0.056    &{0.159}    &0.316   & 0.771   & -       & -     & -     & -  \\
Query-LIFE             & 0.125   & \textbf{0.215}  & {0.340}  & {0.865}  & {0.079}   & \textbf{0.204}  & \textbf{0.329} & \textbf{0.871} & {0.102}   & \textbf{0.215}   & \textbf{0.386}    & \textbf{0.891}  \\
Query-LIFE w/o QMA     & 0.068  & 0.170  & 0.318  & 0.741  & {0.079}  & 0.147   & 0.306   & 0.805  & 0.068   & 0.193   & {0.329}  & 0.784   \\
Query-LIFE w/o QMF     & {0.136}   & 0.207  & 0.318      & 0.856  & 0.090   & 0.147  & 0.306  & {0.863}  & \textbf{0.113}   & {0.193}   & 0.295  & {0.877}   \\
Query-LIFE w/o GenFilt & 0.102       & 0.147     & 0.261     & 0.816     & 0.056      & 0.147      & \textbf{0.329}     & 0.835     & 0.090       & 0.136     & 0.295     & 0.849  \\\bottomrule
\end{tabular}
\end{table*}

\textbf{Online Evaluation Metrics.} We adopt number of orders (Order\_cnt), average number of buyer (Order\_uv), and GMV (Gross Merchandise Volume,total value of sales) as online evaluation metrics. These metrics reflect changes in user order transactions.

\textbf{Human Evaluation.} We sampled 1,000 queries, and selected the top-10 query-item pairs of the exposure page for each query to perform human relevance evaluation. The relevance of a query-item can be divided into three types: Excellent,  Fair and Bad. Excellent means same as the original highly standard, the item's core products, functional attributes and other attributes perfectly match the query requirements.  Fair means  the core product is the same as query, but the functional attributes are inconsistent. Bad means the core products are different or the core products are the same, but the retained attributes in the brand or other key industries are different. We count the proportions of the three indicators of different models.
\subsection{Offline Experiments}
The previous models calculates the cosine similarity between query embedding and title embedding (Query $\rightarrow$ Title) or image embedding (Query $\rightarrow$ Image). In Query-LIFE, we introduce another method which calculates the cosine similarity between  query embedding and $\mathcal{M}$ embedding (Query $\rightarrow$ $\mathcal{M}$).

 AUC of different models can be found in Table~\ref{tab:baseline}.  It is worth noting that the AUC  of general VLP models (CLIP, BLIP2) is lower than that of BERT, because these  general VLP models only focus on the internal alignment of product titles and images, ignoring the imbalance problem between query and products, while BERT  performs external alignment of query and product.  This shows that external alignment improves query and product relevance. In addition, AUC for query$\rightarrow$$\mathcal{M}$ proposed by Query-LIFE is 0.021(0.891-0.871) higher than BERT. It is shown that the relevance score is effectively improved  by introducing image information and external alignment of query-product. 

\begin{figure}[htb]
  \includegraphics[width=0.9\linewidth]{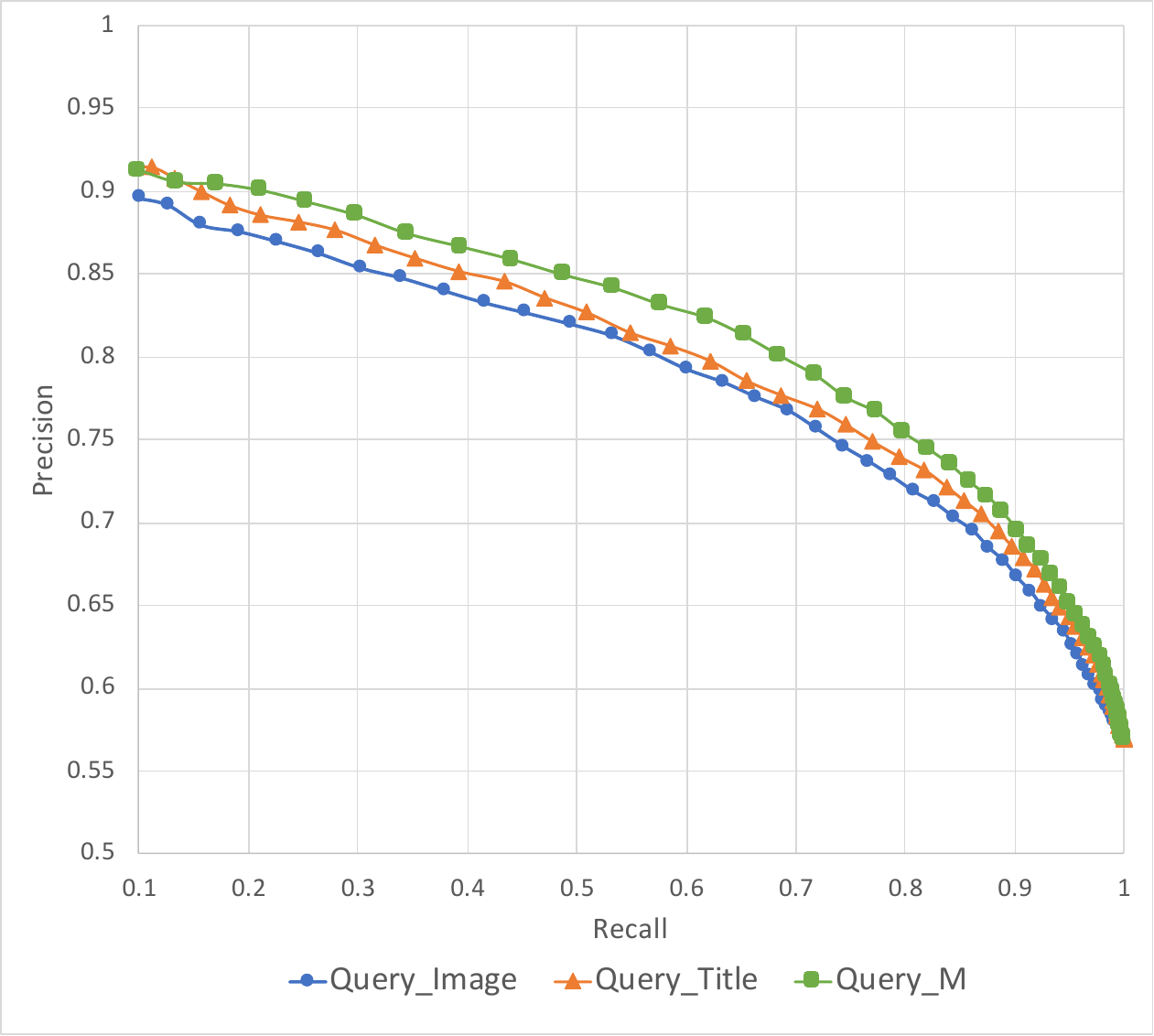}
  \caption{Precision-Recall curve of three  types of representation}
  \Description{.}
  \label{fig:PR}
\end{figure}

Moreover, In Query-LIFE,  when comparing the different relevance score inner products, we observe that the AUC of query $\rightarrow$ $\mathcal{M}$ is also the highest. This indicates that $\mathcal{M}$ representation can provide a more comprehensive and robust representation compared to using a single modality (text or image). To further evaluate these methods, we also plot Precision-Recall (PR) curves of three relevance score inner products in Figure~\ref{fig:PR}. It can be observed that the PR curve of  query $\rightarrow$ $\mathcal{M}$ consistently outperforms the others, indicating the superiority of the query $\rightarrow$ $\mathcal{M}$. 

The R@K results for different models are also presented in Table~\ref{tab:baseline}. Firstly, general VLP model's R@K is lower than BERT and Query-LIFE, further illustrating the impact of query-product external alignment on the relevance score. Secondly,  at R@10 and R@20, query$\rightarrow$$\mathcal{M}$ of  Query-LIFE is 0.029 (0.215-0.186), 0.035 (0.386-0.351) higher than query$\rightarrow$title of BERT.  Query-LIFE enhances the external alignment between the product and query, and effectively enhance $\mathcal{M}$ representation by making full use of image information.

As shown in Figure~\ref{fig:relation}(b), previous VLP models usually adopt divide-and-conquer approach  to utilize image information in e-commece relevance task. They extract query, title, and image representations, and then add query$\rightarrow$title (Q$\rightarrow$T) and query$\rightarrow$image (Q$\rightarrow$I)  as the relevance score. It assumes equal contribution of product text and image for the relevance judgement given any types of products. 
Hence, we compare the performance of  query$\rightarrow$$\mathcal{M}$ and divide-and-conquer approach.  As shown in Table ~\ref{tab:q-m},  query$\rightarrow$$\mathcal{M}$ outperforms the divide-and-conquer approach in all metrics. It clearly demonstrates the advantage of dynamic weighting for different types of products.
\begin{table}[htb]
  \centering
\caption{The R@K and AUC of divide-and-conquer approach and query $\rightarrow$$\mathcal{M}$.}
  \label{tab:q-m}
  \begin{tabular}{cccccc}
  \toprule
                     & Model      & R@5    & R@10   & R@20   & AUC   \\\midrule
  \multirow{2}{*}{$\frac{Q \rightarrow T+ Q \rightarrow I}{2}$} & BLIP2-FT   & 0.090  & 0.181 & 0.329 & 0.781 \\
                     & Query-LIFE & 0.079 & 0.215 & 0.318 & 0.882 \\
Query$\rightarrow$$\mathcal{M}$ & Query-LIFE & 0.102 & 0.215 & 0.386 & 0.891  \\\bottomrule
  \end{tabular}
  \end{table}

\subsection{Online Experiment}
Furthermore, we carry out online A/B experiments, and BERT is the current baseline in Miravia Search. Annotators are invited to evaluate whether the relevance is improved by the proposed method. 10K query-item pairs are sampled top-10 items both from the buckets of BERT and Query-LIFE respectively. The results are shown in Table ~\ref{tab:human}, compared with BERT, the main improvement is that the Excellent ratio increased by 4.42\% and the Bad ratio decreased by 2.79\%. Since the search relevance is one key aspect of user experience, the conversion efficiency is also improved as a consequence. As shown in Table ~\ref{tab:online}, all the efficiency metrics are increased. The results verified that Query-LIFE can attracts higher conversions for our platform. Query-LIFE has been deployed online and brings stable conversion improvements for Miravia Search.


\begin{table}[h]
  \centering
  \caption{Results of human evaluation.}
   \label{tab:human}

\begin{tabular}{cccc}
   \toprule

            & Excellent & Fair & Bad  \\\midrule
Query-LIFE & +4.42\%    & +2.17\%    & -2.79\% \\\bottomrule
\end{tabular}
\end{table}

 \begin{table}[htb]
 \centering
  \caption{Online A/B tests of  Query-LIFE.}
  \label{tab:online}
  \begin{tabular}{cccccc}
    \toprule
 & Order\_cnt & Order\_uv   & GMV  \\\midrule
Query-LIFE  & +4.11\%       & +3.06\%         & +3.19\% \\
  \bottomrule
\end{tabular}
\end{table}

\section{Ablation  Studies}
\subsection{Effect of QMA and QMF} In this section, we will demonstrate what important role QMA and QMF play for Query-LIFE. As listed in Table ~\ref{tab:baseline}, Query-LIFE w/o QMA, Query-LIFE w/o QMF are compared with Query-LIFE. It is seen that QMA significantly improves both R@K and AUC by enhancing the similarity between query and products. Without QMA module, AUC of query$\rightarrow$title, query$\rightarrow$image and query$\rightarrow$$\mathcal{M}$ decreases 0.124 (0.865-0.741), 0.066 (0.871-0.805) and 0.107 (0.891-0.784) respectively. In query$\rightarrow$$\mathcal{M}$,  R@K decreases 0.034 (0.102-0.068), 0.022 (0.215-0.193) and 0.057 (0.386-0.329)  respectively. Because in e-commerce scenarios, product relevance is closely related to user queries,  QMA aligns user queries and different modalities of positive samples through contrastive learning, strengthening relevance score.

QMF is designed to smartly extract information with different weights based on product types. For verification, we sample several different types of products and adopt Query-LIFE to extract text, image and $\mathcal{M}$ representation respectively. Then we calculate the similarity between $\mathcal{M}$ representation and text\&image representation respectively and then normalize them. As shown in Table~\ref{tab:weight}, without QMF, the model assigns equal weights to the image and text representation of different types of products. However, Query-LIFE dynamically adjusts the weights assigned to both image and text to better represent the product. In particular, when it comes to dresses, the model gives more attention to the images than text. Compared with title, image can describe the material, size, pattern and other visual information of the dress more intuitively. Similarly, it also works for other product types.  In addition, without QMF module, AUC of query$\rightarrow$title, query$\rightarrow$image and query$\rightarrow$$\mathcal{M}$ drops 0.009 (0.865-0.856), 0.008 (0.871-0.863) and 0.014 (0.891-0.877) respectively.  Therefore it makes sense by giving more weight to the images. 
\begin{table}[htb]
 \centering
 \caption{The weight of different models on image information.}
  \label{tab:weight}
\begin{tabular}{ccccc}
\toprule
              & Dress  & Monitor   & Phone \\
              \midrule
Query-LIFE        & 0.71  & 0.56     & 0.35  \\
Query-LIFE w/o QMF & 0.49 & 0.52    & 0.51 \\\bottomrule
\end{tabular}
\end{table}

\begin{figure*}[htb]
  \centering
  \begin{subfigure}{0.24\textwidth}
    \includegraphics[width=\linewidth]{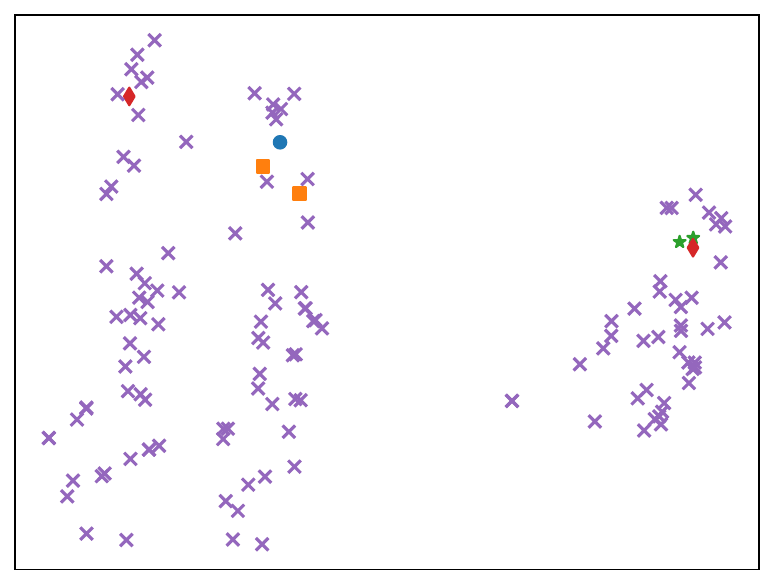}
    \caption{BLIP2-zeroshot}
    \label{fig:BLIP2-zero-shot}
  \end{subfigure}
  \hfill
  \begin{subfigure}{0.24\textwidth}
    \includegraphics[width=\linewidth]{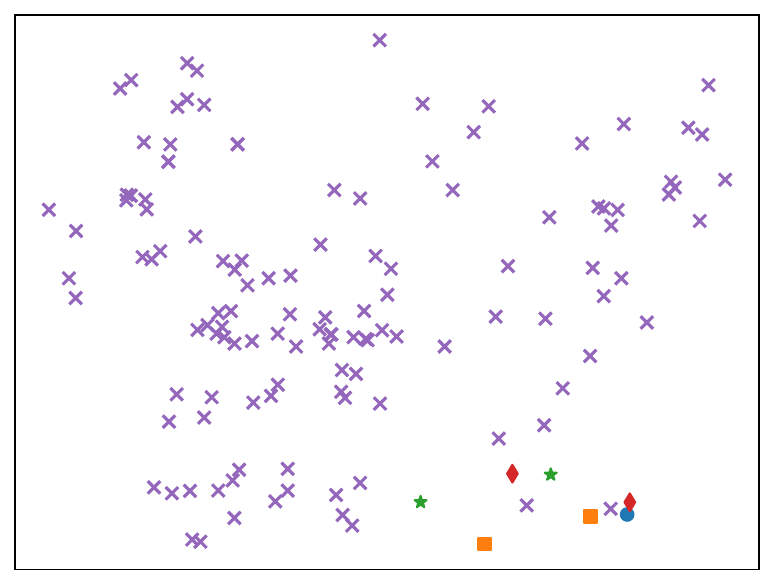}
    \caption{BLIP2-FT}
    \label{fig:itc_query}
  \end{subfigure}
  \hfill
   \begin{subfigure}{0.24\textwidth}
    \includegraphics[width=\linewidth]{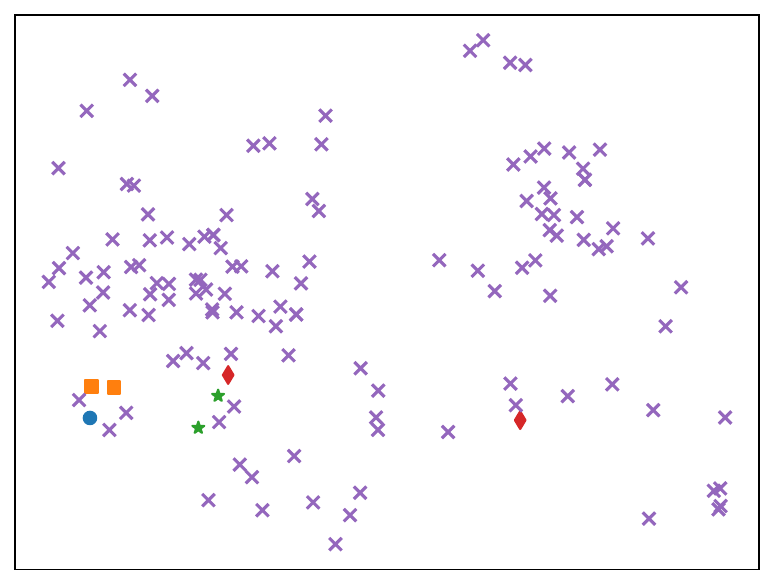}
    \caption{Query-LIFE w/o GenFilt}
    \label{fig:wogenfilt}
  \end{subfigure}
  \hfill
  \begin{subfigure}{0.24\textwidth}
    \includegraphics[width=\linewidth]{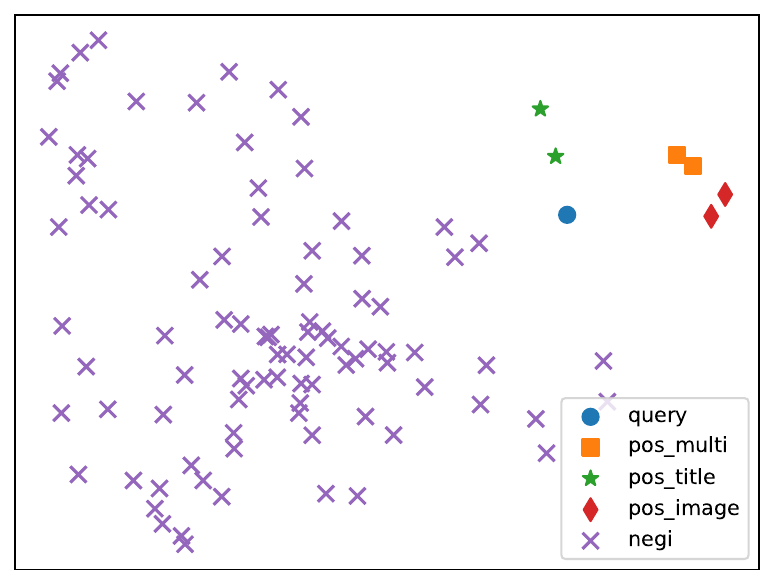}
    \caption{Query-LIFE}
    \label{fig:qlifr}
  \end{subfigure}
  \caption{Embedding space visualization of different models. The purple cross is the different modal representation of the negative samples, the blue dot is the query representation, the orange square is the $\mathcal{M}$ representation of the positive samples, the green five-pointed star is the title representation of the positive samples, and the red diamond is the image representation of the positive samples.}
  \label{fig:embedding}
\end{figure*}

\subsection{Effect of GenFilt}
We fine-tune Query-LIFE w/o GenFilt and list AUC and R@K in Table ~\ref{tab:baseline}. Taking query $\rightarrow$ $\mathcal{M}$ as an example, compared with Query-LIFE, AUC decreases by 0.042 
 (0.891-0.849), R@5, R@10, and R@20 decreases by 0.012 (0.102-0.090), 0.079 (0.215-0.136) and 0.091 (0.386-0.295) respectively. This further shows that in the <query,title,image> data, in-batch sampling will sample some false negative samples and affect model performance. 

We visualize the embedding space of different relevance models (BLIP2-zeroshot, BLIP2-FT, Query-LIFE w/o GenFilt and Query-LIFE) in Figure~\ref{fig:embedding}(a)-(d) respectively. In detail, we sampled positive samples and negative samples for a query. In Figure ~\ref{fig:embedding}(a), the similarity between the query and its positive products does not widen the gap with other negative products. After vision-language pre-training (BLIP2-FT), query and different modals of positive products closer in Figure ~\ref{fig:embedding}(b), while query and some negative samples are still confused.   While GenFilt is not used in Figure~\ref{fig:embedding}(c), the performance is similar with BLIP2-FT in Figure~\ref{fig:embedding}(b). 
 Figure~\ref{fig:embedding}(d) shows the Query-LIFE results, positive samples are more clustered and are further away from negative samples, compared with Figure~\ref{fig:embedding}(c), it intuitively demonstrates the enhancement of model performance by GenFilt.

Finally, we select some cases to show the  generation capability of LLM and InstructBLIP, as shown in Table ~\ref{tab:genfilt},  GenFilt can extract key information such as core products, materials, brand,colors, etc. from titles and images through LLM and InstructBLIP. We adopt these prompts:
\begin{itemize}
    \item \textbf{LLM:} As a product search engine, please understand the input of the product title, extract the core word, material, brand, color, and model parameters from the title and provide structured output.
    The input title: {title}
    To solve the problem, please execute the following steps: Firstly, understand the input product title and extract the vocabulary that describes the main product  as the core word. Secondly, analyze the main material of the product and replace it with "NULL" if none is specified. Thirdly, analyze the brand of the product and replace it with "NULL" if none is specified. Firthly, analyze the color of the product and replace it with "NULL" if not specified. Finally, output the structured parsing results.
    \item \textbf{InstructBLIP:} Briefly summarize the items in the picture in a few words.
\end{itemize}
\begin{table*}[phtb]
  \centering
  \caption{GenFilt Results for some products.}
   \label{tab:genfilt}
  \begin{tabular}{  p{1cm} | p{9cm} | p{3cm} | p{3cm} }
    \toprule
    Image & Title & LLM & InstructBLIP \\ \midrule
    \begin{minipage}[b]{0.13\columnwidth}
		\centering
		\raisebox{-.5\height}{\includegraphics[width=\linewidth]{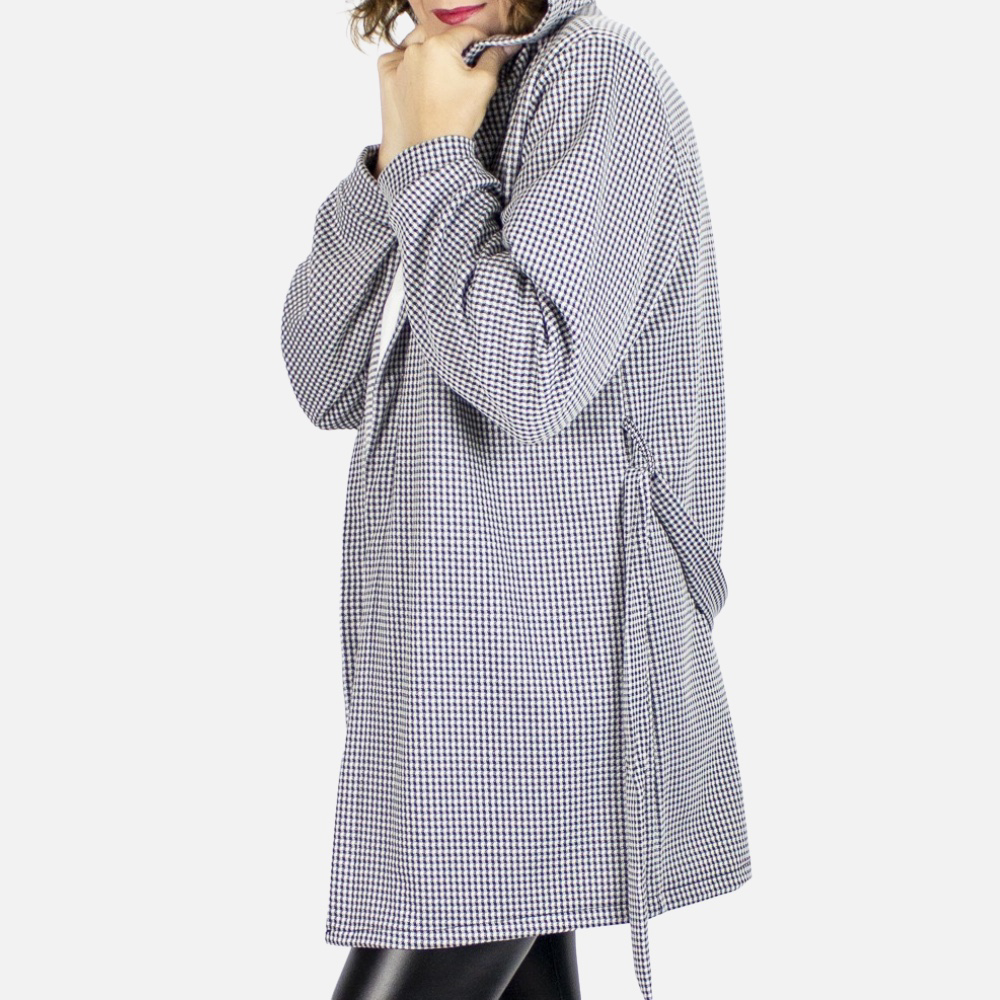}}
	\end{minipage}
    & Women's long jacket with hood houndstooth print black and white casual autumn
    & Women long white jacket
    &Woman white jacket 
    \\ \midrule
    \begin{minipage}[b]{0.13\columnwidth}
		\centering
    \raisebox{-.5\height}{\includegraphics[width=\linewidth]{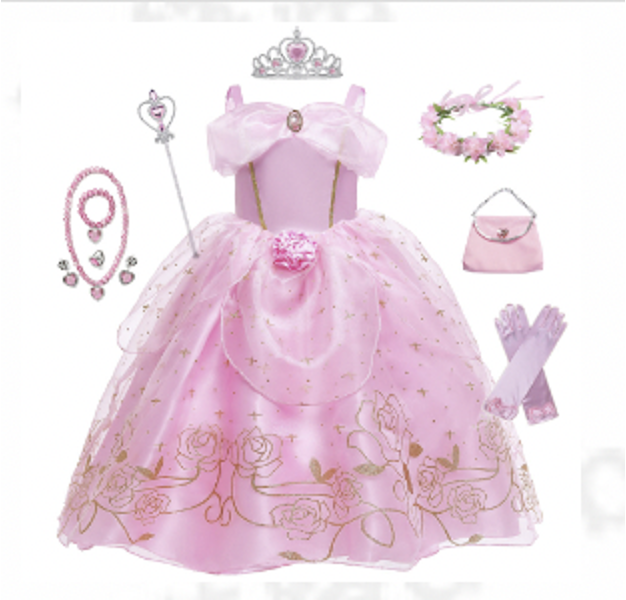}}
	\end{minipage}
    & Kid Princess Dress Girl Summer Fancy Party Clothes Girls Aurora Rapunzel Cinderella Sleeping Beauty Christmas Carnival Costume
    & Dress princess summer fancy party
    & A pink princess dress and accessories
    \\ \midrule
    \begin{minipage}[b]{0.13\columnwidth}
		\centering
    \raisebox{-.5\height}{\includegraphics[width=\linewidth]{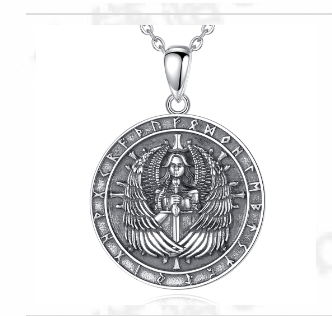}}
	\end{minipage}
    & EUDORA 925 Sterling Silver Viking Odin God Medal Necklace for Man Women Cool Vikings Circle Norse Runes Amulet Pendant Jewelry
    & Necklace silver viking odin god medal
    & A silver pendant with an angel on it
    \\ \midrule
    \begin{minipage}[b]{0.13\columnwidth}
		\centering
    \raisebox{-.5\height}{\includegraphics[width=\linewidth]{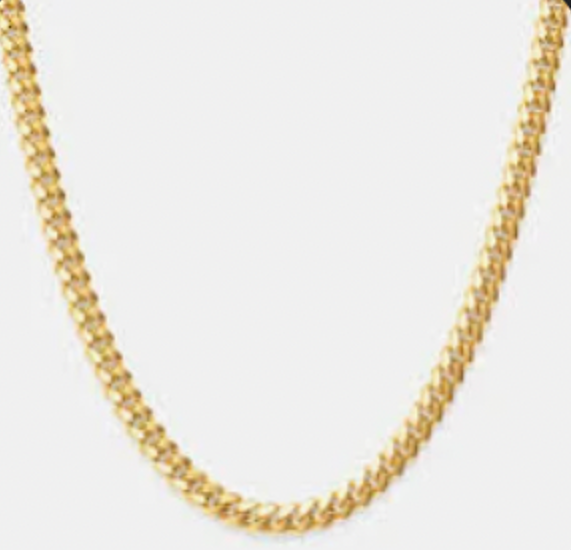}}
	\end{minipage}
    & Gold Blade Chain Choker Necklace Women Sexy Flat Snake Chain Gift
    & Gold necklace blade chain choker
    & A gold chain necklace
    \\\bottomrule
  \end{tabular}
\end{table*}

\section{Conclusion}
In this paper, we propose a novel approach for learning multi-modal representation of product in e-commerce search relevance. We design a query-based multi-modal fusion module which effectively generates dynamic fusion
representations that incorporate product image and text
based on the product types. We propose query-based modal alignment module which utilizes supervised contrastive learning to align the multi-modal representation of products guided by the search query. Additionally, we propose the GenFilt module, leveraging LLM (large language model) and the generation capability extracting information from image to text to address the problem of false negative sampling in contrastive learning. Experimental results demonstrate that Query-LIFE outperforms existing baselines in
both the relevance task. Moreover, Query-
LIFE has been successfully deployed in Miravia Search, leading to improvements both in search relevance and the conversion rate.

\bibliographystyle{ACM-Reference-Format}
\bibliography{sample-base}


\end{document}